\documentclass[useAMS,usenatbib]{aa}

\usepackage{graphicx}
\usepackage{txfonts}
\usepackage{lipsum}
\usepackage{subcaption}         % necessary for continued figures, example in section 3
\usepackage{lscape}             % to rotate a single page table, example in appendix.
\usepackage{placeins}           % useful with \FloatBarrier, to keep 
\usepackage[dvipsnames]{xcolor}
\usepackage{hyperref}

\hypersetup{colorlinks=true, citecolor=MidnightBlue, linkcolor=MidnightBlue}

\newcommand{\ciii}{\hbox{C\,{\sc iii}]}\xspace}
\newcommand{\gsz}{GS-z14\xspace}

\begin{document}

%%%%%%%%%%%%%%%%%%%%%%%%%%%%%%%%%%%%%%%%
% if you use custom commands in your title,
% ensure to check your title when submitting!
%%%%%%%%%%%%%%%%%%%%%%%%%%%%%%%%%%%%%%%%
   \title{Intense and extended CIII] emission suggests a strong outflow in JADES-GS-z14-0}

   %\subtitle{Intense and extended CIII] emission in JADES-GS-z14}

%%%%%%%%%%%%%%%%%%%%%%%%%%%%%%%%%%%%%%%%
% Please separate each author with the \and command
%
% Please do not include ORCIDs next to author names.
% Only ORCIDs authenticated by individual authors in EDPS
% editorial system will be taken into account.
% ORCIDs included here will be removed.
%%%%%%%%%%%%%%%%%%%%%%%%%%%%%%%%%%%%%%%%

   \author{
   Stefano Carniani\inst{\ref{SNS}}\thanks{stefano.carniani@sns.it} \and 
   Peter Jakobsen\inst{\ref{DAWN},\ref{Niels}} \and 
   Giacomo Venturi\inst{\ref{SNS}} \and 
   Francesco D'Eugenio\inst{\ref{Kavli},\ref{Cavendish}} \and 
   Tobias J. Looser\inst{\ref{Harvard}} \and
   Joris Witstok\inst{\ref{DAWN},\ref{Niels}} \and 
   Christopher N. A. Willmer\inst{\ref{Tucson}} \and
   Andrea Ferrara\inst{\ref{SNS}} \and 
   Zihao Wu\inst{\ref{Harvard}} \and
   Santiago Arribas\inst{\ref{CAB}} \and
   Andrew J.~Bunker\inst{\ref{Oxford}} \and   
   Stéphane Charlot\inst{\ref{Paris}} \and 
   Jacopo Chevallard\inst{\ref{Oxford}} \and
   Mirko Curti\inst{\ref{INAFBO}} \and 
   Emma Curtis-Lake\inst{\ref{hert}} \and
   Daniel J.\ Eisenstein\inst{\ref{Harvard}} \and
   Kevin Hainline\inst{\ref{Tucson}}\and
   Jakob M. Helton\inst{\ref{pennsy}}\and
   Zhiyuan Ji\inst{\ref{Tucson}} \and
   Xihan Ji\inst{\ref{Kavli},\ref{Cavendish}} \and  
   Benjamin D.\ Johnson\inst{\ref{Harvard}} \and
   Mahsa Kohandel\inst{\ref{SNS}} \and
   Nimisha Kumari\inst{\ref{aura}} \and
   Roberto Maiolino\inst{\ref{Kavli},\ref{Cavendish},\ref{UCL}} \and
   Andrea Pallottini\inst{\ref{unipi}} \and
   Eleonora Parlanti\inst{\ref{SNS}} \and
   Pablo G. Pérez-González\inst{\ref{CAB}} \and
   Marcia Rieke \inst{\ref{Tucson}}\and
   Pierluigi Rinaldi\inst{\ref{stsci}}\and
   Brant Robertson\inst{\ref{SantaCruz}} \and 
   Jan Scholtz\inst{\ref{Kavli},\ref{Cavendish}} \and  
   Sandro Tacchella\inst{\ref{Kavli},\ref{Cavendish}} \and 
   Hannah {\"U}bler \inst{\ref{MPE}} \and 
   Chris Willot\inst{\ref{Herzberg}} 
    }

   \institute{
    Scuola Normale Superiore, Piazza dei Cavalieri 7, I-56126 Pisa, Italy\label{SNS} \and
    Cosmic Dawn Center (DAWN), Copenhagen, Denmark\label{DAWN} \and
    Niels Bohr Institute, University of Copenhagen, Jagtvej 128, DK-2200, Copenhagen, Denmark \label{Niels} \and  
    Kavli Institute for Cosmology, University of Cambridge, Madingley Road, Cambridge, CB3 0HA, UK\label{Kavli} \and
    Cavendish Laboratory, University of Cambridge, 19 JJ Thomson Avenue, Cambridge, CB3 0HE, UK\label{Cavendish} \and
    Center for Astrophysics $|$ Harvard \& Smithsonian, 60 Garden St., Cambridge MA 02138 USA\label{Harvard} \and  
    Steward Observatory, University of Arizona, 933 N. Cherry Avenue, Tucson, AZ 85721, USA\label{Tucson} \and
    Centro de Astrobiolog\'{\i}a (CAB), CSIC-INTA, Ctra. de Ajalvir km 4, Torrej\'on de Ardoz, E-28850, Madrid, Spain\label{CAB} \and  
    Department of Physics, University of Oxford, Denys Wilkinson Building, Keble Road, Oxford OX13RH, UK\label{Oxford} \and 
    Sorbonne Universit\'e, CNRS, UMR 7095, Institut d'Astrophysique de Paris, 98 bis bd Arago, 75014 Paris, France\label{Paris} \and
    INAF, Osservatorio di Astrofisica e Scienza dello Spazio, Via P. Gobetti 93/3, I-40129 Bologna, Italy \label{INAFBO} \and
      Centre for Astrophysics Research, Department of Physics, Astronomy and Mathematics, University of Hertfordshire, Hatfield AL10 9AB, UK\label{hert} \and
      Department of Astronomy \& Astrophysics, The Pennsylvania State University, University Park, PA 16802, USA\label{pennsy} \and
      AURA for European Space Agency, Space Telescope Science Institute, 3700 San Martin Drive. Baltimore, MD, 21210\label{aura} \and
      Department of Physics and Astronomy, University College London, Gower Street, London WC1E 6BT, UK\label{UCL} \and
      Dipartimento di Fisica “Enrico Fermi,” Universitá di Pisa, Largo Bruno Pontecorvo 3, Pisa I-56127, Italy\label{unipi} \and
      Space Telescope Science Institute, 3700 San Martin Drive, Baltimore, Maryland 21218, USA\label{stsci} \and
    Department of Astronomy and Astrophysics, University of California, Santa Cruz, 1156 High Street, Santa Cruz, CA 95064 USA\label{SantaCruz} \and 
      Max-Planck-Institut f\"ur extraterrestrische Physik, Gie{\ss}enbachstra{\ss}e 1, 85748 Garching, Germany\label{MPE}  \and
    NRC Herzberg, 5071 West Saanich Rd, Victoria, BC V9E 2E7, Canada\label{Herzberg}
    }

   \date{Received September 30, 20XX}

% \abstract{}{}{}{}{}
% 5 {} token are mandatory
 
\abstract{
JWST has revealed an overabundance of very bright, blue galaxies at $z \gtrsim 10$, raising fundamental questions about how star formation and feedback operate at Cosmic Dawn. We present new JWST/NIRSpec MSA PRISM/CLEAR spectroscopy of JADES-GS-z14-0 ($z \simeq 14.18$) obtained with the JADES and OASIS programmes. While the rest-frame UV continuum flux level and shape are consistent between the two datasets, the OASIS spectrum shows a $10\sigma$ detection of the C\,{\sc iii}] $\lambda\lambda1907,1909$ emission line, with a luminosity three times higher than that measured in the JADES data. 
This difference is naturally explained by the offset in shutter placement between OASIS and JADES, implying that the \ciii\ emission is spatially displaced by $\sim400$~pc from the stellar continuum. The non-detection of \ciii\ in ultra-deep NIRCam medium-band imaging indicates that the emitting region is extended on scales $\gtrsim165$ pc, with a surface brightness below the detection threshold. Interpreting this diffuse, carbon-enriched gas as the result of ongoing or past outflows, we infer a mass outflow rate of $\dot{M}_{\rm out}=160~{\rm M_\odot\,yr^{-1}}$. We compare it with the star-formation rate (SFR) and derive a mass-loading factor of $\eta = \dot{M}_{\rm out}/{\rm SFR} = 4-15$, suggesting highly efficient feedback at very early times. Finally, we show that, if outflows are one of the mechanisms regulating star formation in JADES-GS-z14-0, the instantaneous star-formation efficiency in massive haloes is constrained to $\epsilon_\star \lesssim 0.08$.
These results support a scenario in which outflows play a crucial role during the earliest phases of galaxy formation. Comparing our results with the current theoretical galaxy formation model, we conclude that a combination of moderate star-formation efficiency and reduced dust attenuation can account for the emergence of luminous galaxies at the highest redshifts.
}

   \keywords{galaxies: evolution – galaxies: formation – galaxies: high-redshift – galaxies: ISM}

   \titlerunning{Intense and extended CIII] emission in JADES-GS-z14-0}

   \maketitle
    \nolinenumbers

%%%%%%%%%%%%%%%%%%%%%%%%%%%%%%%%%%%%%%%%%%%%%%%%%%%%%%%%%%%%%%
\section{Introduction}

Theoretical models predict that massive galaxies formed within the first $\sim 500$ Myr of cosmic history, at redshifts $z \gtrsim 10$. This epoch, often referred to as the Cosmic Dawn, has long remained largely uncharted territory for observational astrophysics, with only a handful of known galaxy candidates \citep[see review by][]{robertson2022}. In recent years, however, JWST observations with NIRCam and MIRI have led to the discovery of numerous new galaxy candidates at $z \gtrsim 10$ \citep{finkelstein2023, arrabal-haro2023, wang2023, tacchella2023, perez-gonzalez2023, hsiao2024, zavala2024, robertson2024, hainline2024, helton2025}, reported in the literature at a near-monthly cadence and extending to extreme photometric redshifts of $z \sim 16-30$ \citep{castellano2025, perez-gonzalez2025, hainline2026}. Several bright systems at $z \sim 10-15$ have also been spectroscopically confirmed with NIRSpec, showing clear Lyman-break signatures and, in some cases, robust detections of rest-frame UV emission lines \citep{curtis-lake2023, bunker2023, castellano2024, hainline2024, carniani2024, witstok2024a, deugenio2024, witstok2026, naidu2026, donnan2026}.

One of the major puzzles raised by JWST observations concerns the apparent overabundance of very bright ($M_{\rm UV} < -21$) and blue (i.e.\ UV-continuum slope $\beta < -2$) galaxies at $z > 10$, which significantly boosts the bright end of the UV luminosity function relative to pre-JWST expectations \citep[e.g.,][]{finkelstein2023, robertson2023, donnan2024,robertson2024, whitler2025}. The comoving number densities implied by current JWST surveys have proven difficult to reconcile with standard galaxy-formation models calibrated on pre-JWST data. Even when restricting the comparison to spectroscopically confirmed systems, independent studies have reported that the abundance of galaxies at $z \sim 14$ is up to $\sim$100 times higher than expected in the past \citep{harikane2024, robertson2024, whitler2025, naidu2026}. This excess of luminous galaxies has far-reaching implications for our understanding of early galaxy evolution. In particular, it raises fundamental questions about (i) how star formation is regulated \citep{dekel2023, li2024, somerville2025}, (ii) whether stellar mass assembly in the first massive systems is bursty \citep{mason2023, sun2023, pallottini2023, basu2026}, (iii) the presence and properties of dust and its impact on the emergent UV light \citep{ziparo2023, ferrara2023, ferrara2024, ferrara2024c, narayanan2025, sommovigo2026}, (iv) whether the initial mass function (IMF) in the earliest generations of galaxies differed from that inferred at later cosmic times \citep{inayoshi2022, wang2023, trinca2024, yung2024}, (v) wheather the standard $\lambda$CDM picture needs modifications \citep{liu2022, matteri2025}, and (vi) the contribution of accreting massive black holes  \citep{inayoshi2022, trinca2024, hegde2024}.

In the context of the regulation of star formation, two possible scenarios have been proposed. In a weak-feedback scenario, the combination of high gas densities ($n_e>1000~{\rm cm^{-3}}$) and low metallicities, conditions expected in the early Universe, can inhibit the feedback processes regulating star formation. Star-formation timescales become too short for supernova feedback to couple effectively to the interstellar medium \citep{dekel2023, li2024}, while radiation pressure from massive stars acting on the coupled dust--gas medium struggles to overcome the gravitational collapse in molecular clouds with high gas surface densities ($>1000~{\rm M_\odot\,pc^{-2}}$; \citealt{somerville2025}). In such a scenario, star-formation efficiencies are high, producing more massive and hence more luminous galaxies at a fixed halo mass. Conversely, in a standard-feedback scenario, the star-formation efficiency is comparable to that measured in low-$z$ galaxies, and the stellar radiation pressure can be strong enough to drive efficient outflows \citep{fiore2023, ziparo2023, ferrara2023, ferrara2024a, ferrara2024c, ferrara2025}. Such outflows expel gas and dust, and destroy dust grains, reducing the effective dust screen. In this case, the emergence of bright blue galaxies at high redshift is primarily driven by extremely low dust attenuation. Discriminating between these two scenarios requires direct constraints on star formation feedback and efficiency in galaxies at Cosmic Dawn.

Among the known bright galaxies at $z > 10$, JADES-GS-z14-0 (hereafter \gsz) is one of the most intriguing sources. Initially spectroscopically confirmed with JWST/NIRSpec through its Lyman-$\alpha$ break in January 2024 \citep{carniani2024}, the redshift was securely confirmed at $z = 14.1796$ via the detection of [O\,{\sc iii}]88$\mu$m emission with ALMA \citep{schouws2025, scholtz2025a, carniani2025}. NIRCam observations show that this galaxy is significantly extended, with a half-light radius of 260 pc \citep{robertson2024}, indicating that its continuum emission is dominated by stellar light rather than an active galactic nucleus. Recent MIRI observations reported the detection of the rest-frame optical nebular lines [O\,{\sc iii}]$\lambda\lambda4959,5007$ and H$\alpha$, from which a gas-phase metallicity of 10--20\% solar has been derived \citep{helton2025a}. This suggests that a rapid metal enrichment occurred in less than 300 million years after the Big Bang. 

\cite{carniani2025} argue that the combination of strong optical and far-infrared oxygen emission lines and weak \ciii$\lambda\lambda1907,1909$ emission can be only explained by invoking an escape fraction of ionising photons of $11^{+9}_{-7}\%$. This interpretation favours the presence of radiation-driven outflows expelling a substantial fraction of the interstellar medium of the galaxy, hence reducing the gas content and enabling ionising photons to escape. Independent support for this picture came from the non-detection of the ALMA far-infrared continuum, which is also consistent with dust removal or destruction by outflows \citep{schouws2025, schouws2025a, carniani2025}. In addition, the [C~{\sc ii}] $158\,\mu$m line is not detected, implying a low gas reservoir and a gas fraction in the galaxy below 70--80\% \citep{schouws2025a, scholtz2025a}, unusual for a galaxy at such an early epoch. At the same time, the analysis of the prominent damped Ly$\alpha$ absorption indicates the presence of a large amount of diffuse neutral hydrogen on larger scales, just outside the galaxy, potentially deposited there by previous outflow episodes \citep{heintz2025a}.

In this work, we present new NIRSpec PRISM/CLEAR observations from both the JADES (PID 1287) and OASIS (PID 5997) programmes. The new OASIS data reveal an emission line at the expected wavelength of \ciii\ $\sim$3 times brighter than in the JADES spectra, while the flux level and shape of the rest-frame UV continuum are unchanged. Throughout the paper, we investigate the origin of the \ciii\ line and interpret it in the context of early galaxy formation and feedback processes.

Throughout this paper, we adopt the standard cosmological parameters ${\rm H_0 = 70~km s^{-1}~Mpc^{-1}}$, ${\rm \Omega_{\rm M} = 0.30}$, ${\rm \Omega_{\rm \Lambda} = 0.70}$, according to which 1 arcsec at $z = 14.18$ corresponds to a proper distance of 3.237 kpc. Astronomical coordinates and magnitudes are given in the ICRS and AB systems, respectively. We assume an \cite{chabrier2003} IMF with lower and upper mass limits of 0.1 and 300 M$_\odot$, and solar abundances are from \cite{asplund2021}. Finally, as the spectral resolution of the observations is insufficient to resolve the individual lines of the \ciii\ doublet, we refer to the combined emission from the [C~{\sc iii}]$\lambda1906.68\AA$ and C~{\sc iii}]$\lambda1908.73\AA$ transitions simply as \ciii.

%%%%%%%%%%%%%%%%%%%%%%%%%%%%%%%%%%%%%%%%%%%%%%%%%%%%%%%%%%%%%%
\section{Observations}
In this work, we analyse NIRSpec/MOS PRISM/CLEAR observations of \gsz obtained as part of the JADES program (PID 1287; PI: Isaak; \citealt{curtislake2025, scholtz2025, rieke2023, eisenstein2023}) and the OASIS program (PID 5997; PIs: Looser \& D’Eugenio; Looser et al., in preparation). The JADES data (PID 1287) for \gsz\ were acquired by observation 1, visit 1 (1:1) and observation 3, visit 1 (3:1), carried out on 10--11 January 2024 and 12--13 January 2025, respectively, each with 9.3 hours of on-source exposure time. The analysis of the first observation's (1:1) data was presented in \cite{carniani2024}. Here we report for the first time the spectrum from the second observation (3:1). In the OASIS programme (PID 5997), \gsz\ was targeted in a single observation split into two visits (3:1 and 3:2), executed on 5--7 January 2025, between the two JADES observations. The total on-source exposure time of the OASIS data is 27.9 hours.

\begin{figure}
   \centering
   \includegraphics[width=0.9\hsize]{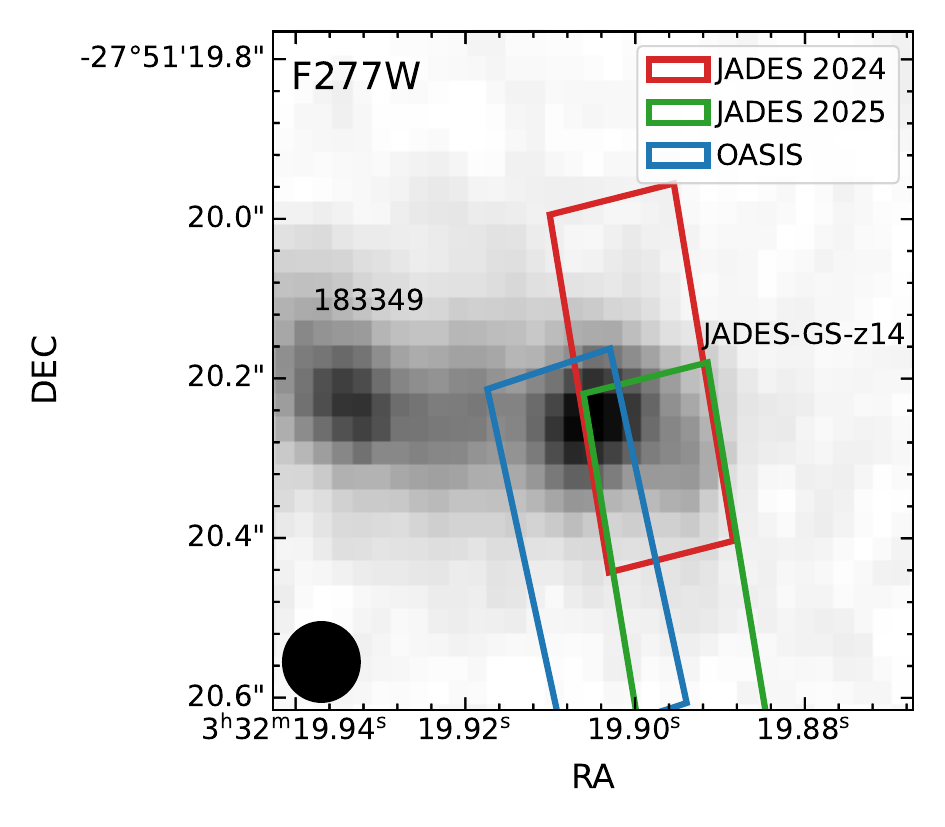}
      \caption{NIRCam F277W image with overlaid the central shutter positions for the two dither visits of JADES (PID 1287; red and green) and the three dithers of OASIS  (PID 5997; blue). The full width at half maximum of the JWST point-spread function at 2.9\,$\mu$m is shown in the bottom-left corner.}
         \label{fig:slit}
   \end{figure}

Figure~\ref{fig:slit} illustrates the slit positions in the different observations and highlights some important differences: the relative placement of \gsz\ within the microshutter changes between the programmes and, for PID 1287, between observations 1:1 and 3:1. These different placements are relevant for both the flux calibration and potential contamination of the extracted spectra. In particular, as discussed below, a neighbouring low-$z$ galaxy may contaminate the OASIS spectrum of \gsz. In the second set of JADES observations (3:1), \gsz is positioned close to the corner of the slit, where  path losses are largest, and the corresponding corrections subject to greater uncertainties \citep{scholtz2025}. 

All NIRSpec data were reduced using version 5.0 of the pipeline developed by the ESA NIRSpec Science Operations Team (SOT) and the NIRSpec GTO team. A detailed description of the workflow is provided in \cite{scholtz2025}. The one-dimensional (1D) spectra were extracted using a refined sigma-clipping algorithm applied to three-pixel apertures (corresponding to $0.3^{\prime\prime}$) from all available subexposures. The pipeline applies a wavelength-dependent slit-loss correction based on the source position within the microshutter, assuming a point-source geometry \citep{scholtz2025}. Because this study aims to compare the spectra from the two programmes in their pipeline-reduced form, we did not renormalise the NIRSpec spectra to the NIRCam flux densities, in contrast to the procedure adopted by \cite{carniani2024}.

\begin{figure*}
   \centering
   \includegraphics[width=0.9\hsize]{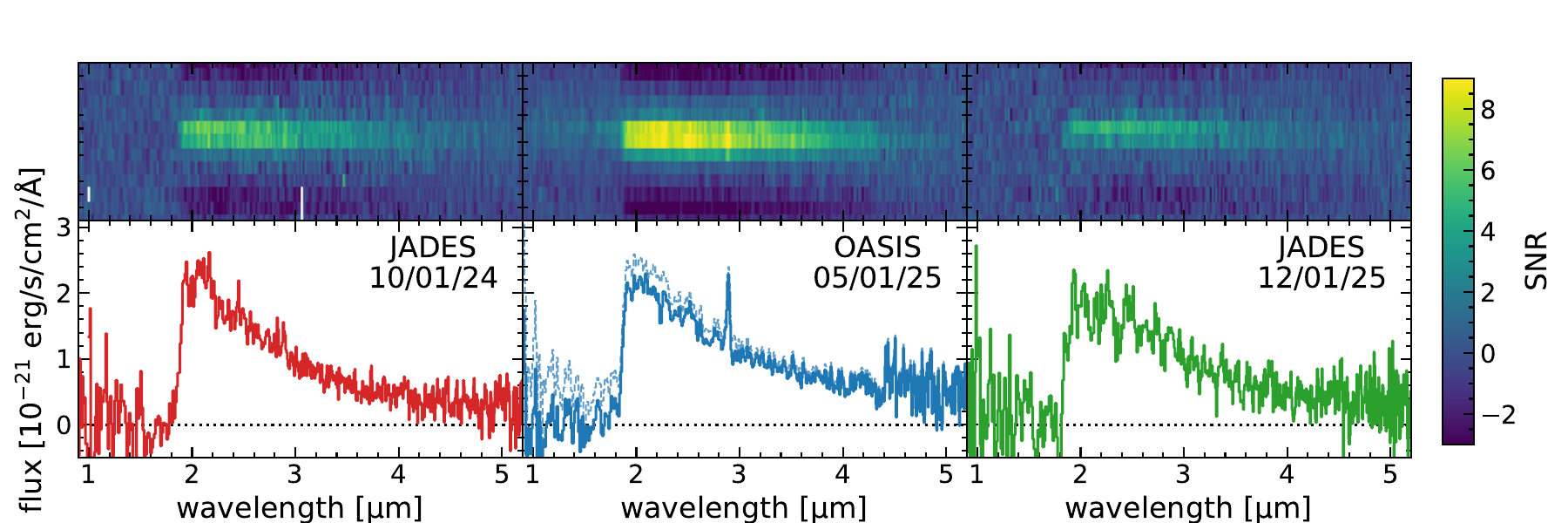}
   \caption{2D SNR maps (top panels) and 1D flux-calibrated spectra of \gsz\ from the JADES and OASIS observations. The data are shown in chronological order adopting the same colour code used for the slits in Fig~\ref{fig:slit}. From left to right, the panels display the 2D and 1D data from the PID 1287 observation obtained in December 2024, the OASIS observation in January 2025, and the second PID 1287 observation conducted one week after the OASIS programme. The dashed blue line in the middle panel reports the 1D spectra obtained from the pipeline, while the solid line illustrates the spectrum after removing the contamination from the neighbouring low-$z$ galaxy.}
         \label{fig:spectra}
   \end{figure*}

Figure~\ref{fig:spectra} shows the one-dimensional (1D) and two-dimensional (2D) spectra of \gsz\ obtained at different epochs from the two programmes. The PID 1287 spectrum collected in 2025 is broadly consistent with that obtained in 2024, although noisier despite an identical exposure time, likely due to increased path losses. In both spectra, the \ciii\ emission line is faint and only marginally detected with a signal-to-noise ratio of 3.6, as reported by \cite{carniani2025}. In contrast, the OASIS spectrum reveals a prominent emission line at the expected wavelength of \ciii\ at $z = 14.18$, with a signal-to-noise ratio of 10. 

%                        A figure as large as the column width
%-------------------------------------------------------------

%-------------------------------------------------------------
%-------------------------------------------------------------

\section{Data analysis of OASIS program}\label{sec:oasis}

In addition to the bright \ciii\ emission line, the OASIS spectrum exhibits several unexpected features for galaxies at $z > 14$. In particular, a distorted continuum profile at wavelengths $\lambda > 4.2\, \mu$m and the presence of non-zero continuum emission shortward of the Ly$\alpha$ limit (see Fig.~\ref{fig:spectra}).

We therefore inspected the background-unsubtracted count-rate maps and identified a diffuse glow at the position of the spectral trace of \gsz, as shown in Fig.~\ref{fig:shorts}. This resembles the artificial signals produced by electronic shorts in the microshutter array circuitry \citep{boker2023, bechtold2025}. This glow is present in all exposures and visits obtained with the same MSA mask configuration, indicating that the artificial emission likely arises from a short associated with one of the selected open shutters. Because the glow contaminates all of the three-shutter slitlets assigned to \gsz, all extracted spectra are affected by this artificial emission. In particular, the spectrum at wavelengths longer than $4.2\, \mu$m is significantly compromised and therefore cannot be reliably used for scientific analysis. 

\begin{figure}
   \centering
   \includegraphics[width=\hsize]{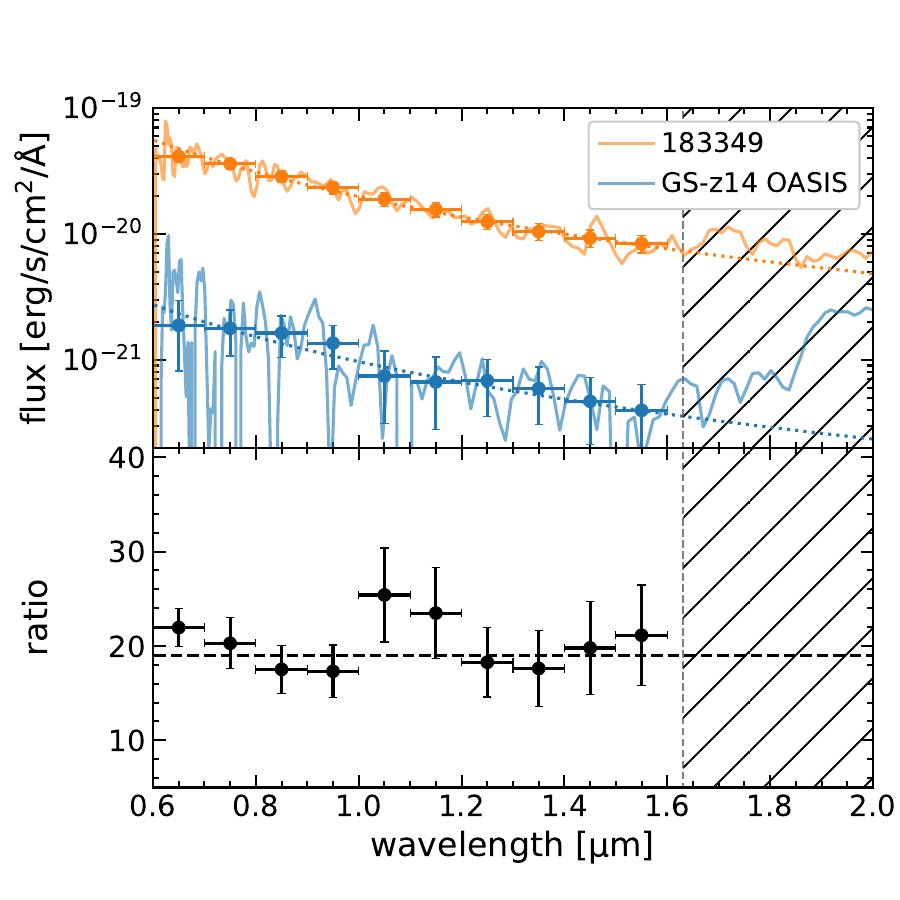}
      \caption{\textit{Top:} Zoom-in of the 1D spectra of GS-z14 from OASIS (blue) and of the galaxy 183349 at $z = 3.47535$ from JADES (orange). Data points with error bars represent the flux densities measured in bins of 0.1~$\mu$m, over which a power-law model ($F_\lambda \propto \lambda^{\beta}$) is fitted. The best-fit models are shown as dotted lines. The grey hatched region indicates the wavelength range excluded from the fit because it lies beyond the Balmer break of the low-$z$ galaxy. \textit{Bottom:} Ratio between the flux densities of galaxy 183349 and the OASIS spectrum. The dashed line shows the best-fit linear model to the data. }
         \label{fig:contamination}
   \end{figure}

We then investigated the systematic emission at wavelengths shortward of Ly$\alpha$. Since the glow produced by microshutter shorts affects the spectrum only at longer wavelengths, we exclude this artefact as the origin of the short-wavelength excess. We compared the continuum emission of \gsz\ in the range $1$--$2\,\mu$m with that of the nearby galaxy 183349 at $z = 3.47535$  (Fig.~\ref{fig:slit}), whose spectrum was obtained in the JADES program \citep{carniani2024}. Figure~\ref{fig:contamination} shows the continuum emission in the two spectra. Wavelengths beyond $1.63\,\mu$m are masked because the comparison is complicated by the Balmer break in the low-$z$ galaxy and the Lyman break in \gsz. By fitting the spectra between $0.6\,\mu$m and $1.63\,\mu$m with a power-law model, $f_\lambda \propto \lambda^{\beta}$, we found $\beta = -2.03 \pm 0.06$ for 183349 and $\beta = -2.0 \pm 0.2$ for the OASIS spectrum. This similarity suggests that the systematic positive emission detected blueward of Ly$\alpha$ in the OASIS data of \gsz\ is likely due to contamination from galaxy 183349. The ratio between the two spectra is shown in the bottom panel of Fig.~\ref{fig:contamination} and is consistent with a constant value of $19.6 \pm 1.1$.

%-------------------------------------------------------------
%-------------------------------------------------------------

\begin{figure*}
   \centering
   \includegraphics[width=0.9\hsize]{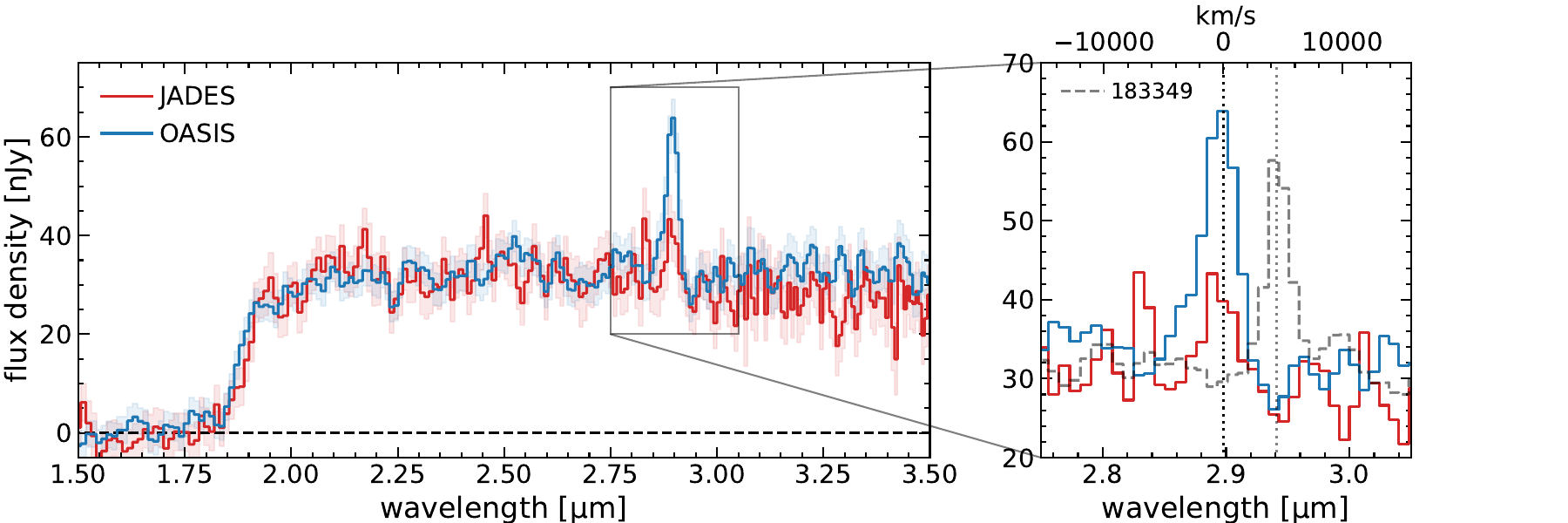}
   \caption{
Comparison of the flux-calibrated NIRSpec spectra of \gsz\ from the JADES (red; \citealt{carniani2024}) and OASIS (blue) observations. 
\textit{Left:} Continuum emission over the wavelength range $1.5$--$3.5\,\mu$m, showing excellent agreement between the two datasets after correcting the OASIS spectrum for contamination from the nearby foreground galaxy continuum, indicating consistent stellar continuum measurements for \gsz across the two different epochs. 
\textit{Right:} Zoom-in around the \ciii\ emission line at $\lambda \simeq 2.895\,\mu$m, shown in both wavelength and velocity space relative to the systemic redshift. 
While the continuum levels are consistent, the OASIS data reveal a significantly stronger \ciii\ emission compared to JADES. 
The grey dashed line indicates the expected wavelength of H$\alpha$ emission from the foreground galaxy 183349 at $z=3.47535$. 
The large velocity offset ($\sim 4000~\mathrm{km\,s^{-1}}$) between its wavelength and that of the detected line rules out contamination from the low-redshift interloper, supporting a scenario in which the \ciii\ emission of \gsz is spatially offset from its stellar component.
}
         \label{fig:comparison}
   \end{figure*}

After subtracting the spectrum of galaxy 183349, rescaled by a factor of 0.051 (i.e., $1/19.6$), from the OASIS spectrum of \gsz, we found the emission at wavelengths shortward of Ly$\alpha$ to be consistent with zero (solid blue line in Fig.~\ref{fig:spectra}). Moreover, the resulting continuum profile is consistent with the spectrum previously analysed in \cite{carniani2024}\footnote{We note that the comparison with the new PID 1287 data is not meaningful, as this dataset cannot be reliably flux-calibrated owing to the large uncertainties in the path-loss correction at the target's position during that observation.}, as shown in Fig.~\ref{fig:comparison}. The agreement between the contamination-corrected OASIS spectrum and the previous JADES data shows that the stellar continuum emission is consistent among the observations. The only significant difference is the presence of a strong carbon emission line in the OASIS data, whose origin is examined in the following section.

%%%%%%%%%%%%%%%%%%%%%%%%%%%%%%%%%%%%%%%%%%%%%%%%%%%%%%%%%%%%%%
\section{The carbon emission line}

%-------------------------------------------------------------
%-------------------------------------------------------------
The OASIS spectrum shows the presence of an emission line at $\lambda = 2.8950 \pm 0.0013~\mu$m with an integrated significance of $10\sigma$. This wavelength is consistent with the tentative detection of \ciii\ reported by \cite{carniani2024} and corresponds to the expected \ciii\ line at $z = 14.1796$ \citep{carniani2025, schouws2025a}. The line is spectrally unresolved, and no kinematic information can be inferred from the observations owing to the low spectral resolution of the PRISM/CLEAR configuration ($\Delta v_{\rm PRISM}\sim3,000~{\rm km~s^{-1}}$ at 3~$\mu$m). The emission line has an integrated flux of $(3.83 \pm 0.38) \times 10^{-19}~{\rm erg~s^{-1}~cm^{-2}}$, corresponding to a luminosity of $L_{\rm CIII]} = (1.13 \pm 0.13) \times 10^{42}~{\rm erg~s^{-1}}$. The measured flux is approximately 2.7 times higher than that derived from the JADES spectrum presented in \citep{carniani2024}. We note that the observed $L_{\rm CIII]}$ is consistent with the predictions expected from SERRA simulations \citep{kohandel2025}.

The bright \ciii\ emission line detected in the OASIS data raises questions about its origin, given its absence in the JADES spectrum. We evaluate three possible scenarios: (1) temporal variability in the line intensity; (2) contamination from a low-redshift interloper; and (3) spatially extended and/or spatially offset emission, which ultimately emerges as the most plausible interpretation.

\subsection{Variability}
If \gsz\ hosts an active galactic nucleus (AGN) at its centre, intrinsic AGN variability could, in principle, account for the observed variation in the \ciii\ line intensity between different epochs \citep[e.g.,][]{trevese2014}. We test this scenario by exploiting the fact that the OASIS observations were obtained approximately one week before the second JADES observation, from which we infer a variability timescale of $\Delta \tau = 7/(1+z)\,{\rm days} = 0.46\,{\rm day}$ in the rest frame system. 

The observed amplitude variation of the \ciii\ emission line, $\Delta m_{\rm CIII]} = 1.1~{\rm mag}$, therefore occurs on a timescale that is too short to be attributed to AGN variability.  Even in extreme cases such as changing-look AGN \citep{ricci2023}, where broad emission lines can appear or disappear and vary in intensity by up to $\Delta m_{\rm BLR-line}\sim1-2$~mag, the characteristic timescales are typically of the order of months to decades \citep{denney2014,zeltyn2024, jana2025}. Therefore, explaining the discrepancy between the spectra solely in terms of AGN-driven variability appears highly unlikely.

\subsection{Contamination}

Based on the contamination discussed in Sect.~\ref{sec:oasis}, we also consider the possibility that the emission line detected in the OASIS data arises from H$\alpha$ emission originating in the neighbouring galaxy 183349 at $z = 3.475$. In this scenario, the line would be observed at $\lambda = 2.9376\,\mu$m, corresponding to a velocity offset of $\sim 4000~{\rm km~s^{-1}}$ relative to the wavelength at which the emission line is actually detected (see Fig.~\ref{fig:comparison}).  Additionally, we would expect to detect the [O\,{\sc iii}]$\lambda5007$ line at $\sim2.24\,\mu$m. The spectral offset, together with the absence of other emission lines, rules out the possibility that the observed \ciii\ emission arises from contamination by 183349.

Another possibility is that the detected emission line originates from H$\alpha$ emission from an interloper at $z = 3.41$. In this case, the inferred H$\alpha$ luminosity would be $L_{\mathrm{H}\alpha} = 4 \times 10^{40}~{\rm erg\,s^{-1}}$, corresponding to a star-formation rate of ${\rm SFR} \approx 0.3~M_\odot\,{\rm yr^{-1}}$ and a rest-frame UV luminosity of $L_{\mathrm{UV}} = 3 \times 10^{42}~{\rm erg\,s^{-1}}$. A galaxy with this luminosity at $z = 3.41$ would be detected in the deep NIRCam F070W and F115W JADES imaging at a significance greater than $5\sigma$ \citep{robertson2024, simmonds2025}. This indicates that identifying the detected line with an optical rest-frame emission line from a low-$z$ galaxy is unlikely.

\subsection{Spatially offset and diffuse emission}\label{sec:offset}

The final possible scenario is that a significant fraction of the \ciii\ emission originates from a region spatially offset with respect to the UV-continuum counterpart. This interpretation is consistent with the fact that the OASIS shutter sampled a portion of the sky different from that covered by the JADES program. While the similar UV-continuum profiles indicate that the stellar component is recovered consistently, approximately two-thirds of the \ciii\ line flux appears to arise from a region located to the east of the galaxy and that was not included in the JADES microshutter configurations (see Fig.~\ref{fig:slit}). The emitting region is therefore constrained to lie between the left edge of the JADES microshutter and that of the OASIS microshutter, as illustrated in Figs.~\ref{fig:slit} and \ref{fig:res}.  Given the size of the microshutter along the dispersion direction (i.e.\ $0.2^{\prime\prime}$; \citealt{ferruit2022, jakobsen2022}), we estimate that the emission originates at a projected distance of $d \approx 400 \pm 150$~pc from the galaxy centre, larger than the stellar scale radius ($R_\star = 260$~pc; \citealt{robertson2024, carniani2024}).

We have also analysed the deep NIRCam F300M image and compared it with other medium- and wide-band filters. No emission feature is detected around the galaxy at a significance greater than $3\sigma$ (Fig.~\ref{fig:res}). To interpret this, we simulated the F300M observations using the JWST Exposure Time Calculator (ETC) v5.1. The simulations show that the displaced \ciii\ emission would have been detectable in the deep JADES NIRCam data only if the emission originated from a compact, point-like source, assuming a $5\sigma$ sensitivity of 1.8 nJy within a $0.1^{\prime\prime}$ aperture \citep{robertson2024}. If instead the emitting region is larger than $0.05^{\prime\prime}$ (i.e.\ $\sim 165$ pc), the surface brightness is sufficiently low that the signal falls below the detection threshold. This result supports a scenario in which the \ciii\ emission arises from diffuse gas spatially offset from the stellar component of the galaxy.

%%%%%%%%%%%%%%%%%%%%%%%%%%%%%%%%%%%%%%%%%%%%%%%%%%%%%%%%%%%%%%
\section{Gas Mass}\label{sec:mass}

The luminosity of the \ciii emission line allows us to estimate the mass of ionised gas traced by this ultraviolet transition, under a set of fiducial assumptions. In this section, we describe in detail the procedure used to derive the gas mass from the \ciii line luminosity.

Following \cite{carniani2015}, the \ciii line luminosity can be expressed as

\begin{equation}
L_{\rm C\textsc{iii}]} = \int_V f n_e n(C^{2+}) j_{\rm C\textsc{iii}]}(n_e, T_e)dV,
\end{equation}
where $f$ is the filling factor, $n_e$ the electron density, $n(C^{2+})$ is the density of $C^{2+}$ ions, and $j_{\rm C\textsc{iii}]}(n_e, T_e)$ is the sum of the 
line emissivities of [C~{\sc iii}]$\lambda1906.68\AA$ and C~{\sc iii}]$\lambda1908.73\AA$ transitions. The density of doubly ionized carbon can be rewritten as
\begin{equation}
n(C^{2+}) = \left(\frac{n(C^{2+})}{n(C)}\right)\left(\frac{n(C)}{n(O)}\right)\left(\frac{n(O)}{n(H)}\right)\left(\frac{n(H)}{n_e}\right) n_e
\end{equation}

where $n(\mathrm{C})$, $n(\mathrm{O})$, and $n(\mathrm{H})$ denote the number densities of carbon, oxygen, and hydrogen, respectively. Assuming $n(\mathrm{H}) \simeq n_e$, the density of doubly ionized carbon ions can be expressed in terms of the carbon-to-oxygen abundance ratio $[\mathrm{C}/\mathrm{O}]$\footnote{We adopt the notation $[\mathrm{X}/\mathrm{Y}] = \log\!\left[n(\mathrm{X})/n(\mathrm{Y})\right] - \log\left[n(\mathrm{X})/n(\mathrm{Y})\right]_\odot$, where  $\log\left[n(\mathrm{C})/n(\mathrm{O})\right]_\odot=-0.26$, and  $\log\left[n(\mathrm{O})/n(\mathrm{H})\right]_\odot=-3.31$.} and the oxygen abundance $[\mathrm{O}/\mathrm{H}]$, both relative to their solar values, as follows: 

\begin{equation}\label{eq:lum}
n(C^{2+}) = 2.69\times10^{-4} f_{\mathrm{C}^{2+}}10^{[C/O]}10^{[O/H]}n_e \,,
\end{equation}
where $f_{\mathrm{C}^{2+}} \equiv n(\mathrm{C}^{2+})/n(\mathrm{C})$ denotes the fraction of carbon in the doubly ionized state.

The line luminosity of the \ciii doublet can be thus rewritten as 
\begin{equation}
L_{\rm C\textsc{iii}]} = 2.69\times10^{-4} f_{\mathrm{C}^{2+}}10^{[C/O]}10^{[O/H]}\langle n_e^2 \rangle j_{\rm C\textsc{iii}]}(n_e, T_e) V \,,
\end{equation}
where $\langle n_e^2 \rangle$ is the volume-averaged square density. The mass of the ionised gas is
\begin{equation}\label{eq:mass}
M_{\rm ion} = \int_V f \bar{m}  n(H) dV \simeq f\, m_p \langle n_e \rangle V\, ,
\end{equation}
where $\bar{m}$ is the average molecular weight within the volume $V$. To first order, we approximate $\bar{m} =1.33m_p$ \citep{carniani2015} and $n(\mathrm{H}) \simeq n_e$ for fully ionized gas\footnote{Any neutral gas entrained in the outflow would further increase the total gas mass, as it would imply $n({\rm H}) > n_e$.}.

Combining the equation \ref{eq:lum} and  \ref{eq:mass}, we obtain
$$
M_{\mathrm{ion}} = \frac{1.33 m_p  C }{ 2.69\times10^{-4} f_{\mathrm{C}^{2+}}10^{[C/O]}10^{[O/H]} j_{\rm C\textsc{iii}]}} \, ,
$$
where $C = \langle n_e \rangle / \langle n_e^2 \rangle$ is the clumping factor.
Assuming a  temperature of  $T_e=10^4K$ and electron density $n_e$= 100 cm$^{-3}$ \citep[e.g.][]{osterbrock2006}, the gas mass is given by
$$
M_{\mathrm{ion}} = 2.12\times10^{6}~{\rm M_\odot} \left(\frac{L_{\rm C\textsc{iii}]}}{10^{40}{\rm erg/s}}\right) \left(\frac{100{\rm cm^{-3}}}{n_e}\right)\frac{C}{ f_{\mathrm{C}^{2+}}10^{[C/O]}10^{[O/H]}} \,.
$$

Although the physical conditions of the displaced gas are uncertain, we estimate the ionised-gas mass by adopting the interstellar medium properties inferred from optical nebular lines \citep{helton2025a}, and assuming that all carbon atoms are doubly ionised ($f_{\mathrm{C}^{2+}} = 1$) and a clumping factor of $C = 1$. Under these assumptions, we obtain an ionised-gas mass of $M_{\rm ion} = 10^{8.5 \pm 0.5}~{\rm M_\odot}$.

%%%%%%%%%%%%%%%%%%%%%%%%%%%%%%%%%%%%%%%%%%%%%%%%%%%%%%%%%%%%%%
\section{CIII] halo}

Spatially displaced and extended carbon emission in primeval galaxies can originate from strong stellar feedback, which numerous theoretical models predict to play a major role in galaxy formation in the early Universe 
\citep[e.g.,][]{salvadori2009, dayal2014, ferrara2023, ferrara2024, mcclymont2025}. In recent years, ALMA observing programmes have reported several detections of spatially offset and extended halo [C\,{\sc ii}] $\lambda158\,\mu$m emission in galaxies at $z > 5$ \citep{carniani2017, carniani2018,carniani2020, matthee2017, matthee2019, fujimoto2019, fujimoto2020}. These detections have often been interpreted as evidence for ongoing outflows, or for large-scale metal-enriched gas distributed by previous feedback episodes \citep{pizzati2020}. Evidence for spatially extended \ciii\ emission, however, is rare, and has so far been reported only for GN-z11 \citep{maiolino2024b}. In that case, the emission was associated with an AGN ionisation cone. Owing to the relatively high ionisation potential required to produce doubly ionised carbon (C$^{2+}$; 24.38 eV), extended \ciii\ emission requires either in-situ excitation or the escape of sufficiently energetic UV photons from the galaxy to ionise carbon in the surrounding medium.

For \gsz, the presence of outflows has already been proposed from theoretical studies to explain its relatively low dust attenuation ($A_{\rm v}\sim0.2$) given its stellar mass ($M_{\rm \star}\sim10^9~{\rm M_\odot}$) and size ($R_\star\sim260$\, pc), and to reproduce its inferred star-formation history \citep{ferrara2024, kohandel2025}. In particular, \cite{ferrara2024} predicted that following an initial phase in which galaxies are heavily dust-obscured, star-formation activity would drive powerful radiation-driven outflows expelling gas and dust. This process removes dust and hence reduces attenuation, making galaxies appear as bright, ``blue'' sources, while temporarily suppressing star formation. A declining star-formation history is further supported by recent MIRI observations of GS-z14 \citep{helton2025a}, which revealed H$\alpha$ emission corresponding to ${\rm SFR_{\rm H\alpha}} = 9.6\pm2.2~{\rm M_\odot\,yr^{-1}}$. This is a factor of 2.6 lower than the ${\rm SFR_{\rm SED}}= 25\pm5~{\rm M_\odot~yr^{-1}}$ inferred from spectral energy distribution fitting, which traces star formation on longer time scales ($\gtrsim10$~Myr). This discrepancy is compatible with a star-formation activity that has declined over the past $\sim 10$ Myr  \citep{helton2025a}.

Outflows are also invoked by \cite{carniani2025} to explain the low gas content and the non-zero escape fraction of ionising photons in \gsz. This scenario is particularly relevant for interpreting the relatively weak \ciii\ equivalent width observed in the JADES data compared to the strength of the [O~{\sc iii}]$\lambda88\mu$m line. The detection of the far-infrared [O~{\sc iii}]$\lambda88\mu$m line implies a gas-phase metallicity of $\sim$10--20\% of the solar value, an estimate further supported by the MIRI observations \citep{helton2025a}. At such metallicities, photoionisation models predict \ciii\ emission approximately 2--3 times stronger than observed, suggesting that additional mechanisms, such as the escape of ionising photons, are required to suppress the UV line intensity
\footnote{Because the production of O$^{2+}$ requires a higher ionisation potential than that of C$^{2+}$, [O~{\sc iii}] emission line traces H\,{\sc ii} regions closer to the surface illuminated by the ionising source than \ciii. Therefore, in a density-bounded configuration, where the outer zones of the H\,{\sc ii} region are truncated, the \ciii flux may be suppressed \citep{plat2019, ramambason2022}, implying a higher [O~{\sc iii}]/\ciii ratio as observed in \gsz.}.

Current observations are insufficient to constrain the origin of the extended \ciii\ emission in \gsz. However, motivated by the theoretical model by \cite{ferrara2024} and the decreasing star-formation history in the last 10 Myr inferred from the JWST data \cite{carniani2025, helton2024} , we speculate that the emission can arise from a \ciii\ halo associated either with an ongoing outflow or with the remnants of past outflows which enriched the circumgalactic medium with carbon. In both scenarios, the absence of rest-frame UV emission in situ from the NIRCam imaging suggests that the extended \ciii\ emission is not powered by local star formation. Instead, it is likely produced by excitation from UV radiation escaping from the central galaxy as shown in Appendix~\ref{sec:photoionization}.

Assuming an ongoing outflow, we estimate the outflow energetics and compare it with the SFR of the galaxy. The mass outflow rate can be approximated as $\dot{M}_{\rm out} = M_{\rm out} v_{\rm out}/R_{\rm out}$, where $M_{\rm out}$, $v_{\rm out}$, and $R_{\rm out}$ denote the outflow mass, velocity, and spatial extent, respectively. The outflow mass is inferred from the \ciii\ luminosity as described in Sect.~\ref{sec:mass}, assuming that all the outflowing gas is ionised ($M_{\rm out} = M_{\rm ion}$). The size can be constrained by noting that the \ciii\ line is detected in the OASIS spectrum, but not in the JADES data (see Sect.~\ref{sec:offset}). This implies a size of $R_{\rm out}= 400$~pc .

The PRISM/CLEAR spectra do not have sufficient spectral resolution to constrain the gas kinematics, and thus, the outflow velocity. However, if the gas has reached a distance of $\sim 400$ pc from the galaxy centre, a conservative lower limit on the velocity is given by the minimum velocity necessary to reach such a distance
$$
v_{\rm out} \approx \sqrt{2GM_{\rm dyn}\left(\frac{1}{R_\star}-\frac{1}{R_{\rm out}}\right)} =170~{\rm km\,s^{-1}},
$$
where we use the dynamical mass estimated by \cite{scholtz2025a}, of $M_{\rm dyn}=10^{9.4}~{\rm M_\odot}$, and the scale radius of the galaxy from \cite{robertson2024} and \cite{carniani2024}, of $R_\star=260~{\rm pc}$. Combining these constraints, we obtain $\dot{M}_{\rm out} = 10^{2.2\pm0.5}~{\rm M_\odot\,yr^{-1}}$, which is 15 times higher than the SFR inferred from H$\alpha$ \citep{helton2025a} and about a factor of 6 higher than the SFR inferred from SED fitting \citep{helton2025, carniani2025}.

Alternatively, if the extended emission represents the remnant of past outflows, we can estimate the average mass outflow rate by assuming that the gas accumulated over the past $\sim 10$ Myr, corresponding to the epoch during which \gsz\ reached the peak of its star-formation activity \citep{carniani2025, ferrara2024c}. Under this assumption, we derive an average mass outflow rate of $\sim 10^2~{\rm M_\odot\,yr^{-1}}$, that is 4 times higher than ${\rm SFR}_{\rm SED}$.

The kinetic energy and momentum rates of the outflow are $\dot E = \frac{1}{2}\dot{M}_{\rm out} v_{\rm out}^2 \sim 9\times10^{41}~{\rm erg\,s^{-1}}$ and $\dot P = \dot{M}_{\rm out} v_{\rm out} \sim 10^{35}~{\rm g\,cm\,s^{-2}}$, respectively. The kinetic energy rate is lower than the energy injection rate expected from supernova feedback, estimated as $\dot E^{\rm SN}_{\rm kin} \sim 7\times10^{41}~{\rm erg\,s^{-1}}\,({\rm SFR}/{\rm M_\odot\,yr^{-1}})$ \citep{leitherer1999, veilleux2005}, corresponding to $\sim 6\times10^{42}~{\rm erg\,s^{-1}}$ when adopting the SFR inferred from the H$\alpha$ luminosity. The momentum rate is comparable to the radiation momentum rate, $\dot P^{\rm rad} = L_{\rm UV}/c \sim 10^{35}~{\rm g\,cm\,s^{-2}}$. This suggests that both supernova feedback and radiation pressure may contribute to driving the outflows observed in \gsz.

\section{Star formation efficiency}

\begin{figure}
   \centering
   \includegraphics[width=\hsize]{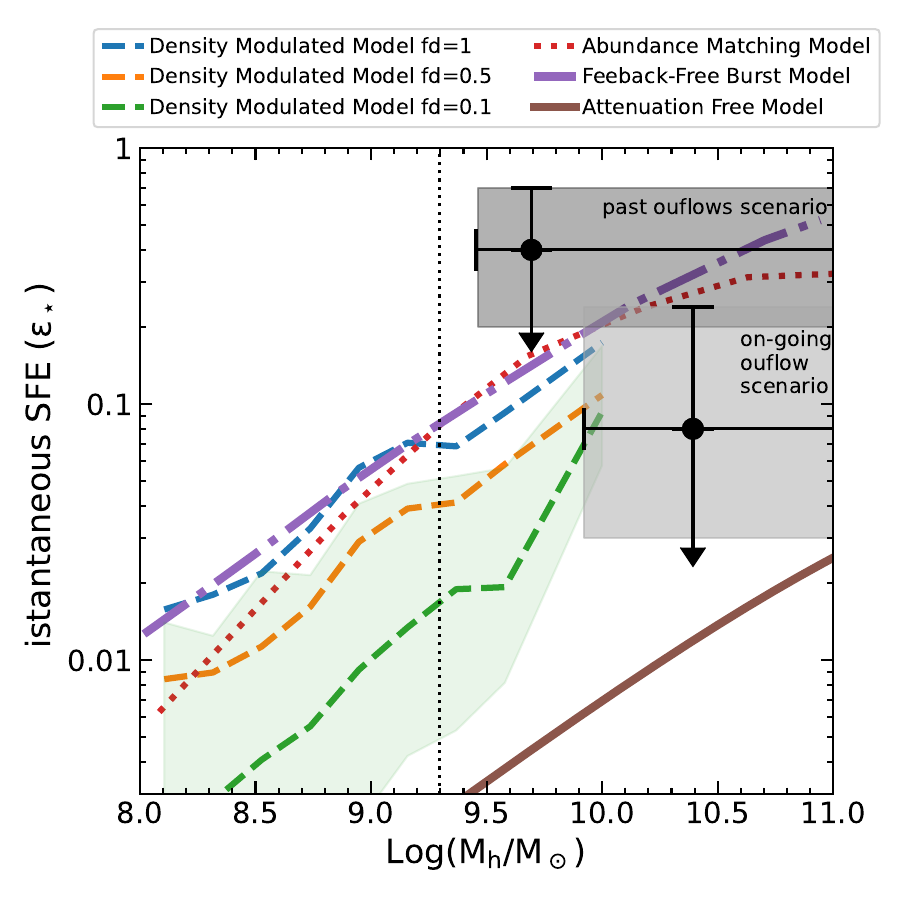}
      \caption{Instantaneous star-formation efficiency ($\epsilon_\star$) as a function of halo mass M$_{\rm h}$. Blue, orange, and green lines illustrate the density-modulated SFE model by \cite{somerville2025} for three different values of the gas fraction in dense clouds, ${\rm f_d}= 1, 0.5, 0.1$. The shaded green region shows the 16th and 84th percentiles of the density-modulated SFE model with ${\rm f_d}= 0.1$. The red line is the result of an empirical model based on abundance matching from \cite{yung2025}. The purple and brown lines illustrate the SFE adopted in the Feeback-Free Burst model \citep{dekel2023, li2024} and Attenuation Free model \citep{ferrara2023, ferrara2024a}. The dashed vertical line reports the minimum value for the halo mass of \gsz, considering the cosmological baryon fraction $f_b = \Omega_b/\Omega_m = 0.158$ and a 100\% star-formation efficiency. The upper limits with the uncertainties for the past outflow scenario and the on-going outflow interpretation are reported as black marks.  The shaded gray regions illustrate the limits on the halo mass for \gsz using the SFE estimated in the two scenarios: M$_{\rm h}=\epsilon_\star f_b M_\star$.}\label{fig:sfe}
   \end{figure}

In summary, under both interpretations, ongoing outflows and remnant gas from past outflows, we infer that a substantial fraction of the gas is expelled to large distances on short timescales, implying a mass-loading factor of $\eta = \dot{M}_{\rm out}/{\rm SFR} \simeq 4-15$ depending on the adopted scenario. Such large mass-loading factors suggest that outflows play a significant role in regulating the star-formation activity of \gsz. If ejective feedback is the dominant regulatory mechanism, we can relate the mass-loading factor to the instantaneous star-formation efficiency, $\epsilon_\star$, by writing the instantaneous SFR as ${\rm SFR} = \epsilon_\star M_{\rm gas}/t_{\rm ff}$, where $M_{\rm gas}$ is the total gas mass in the halo and $t_{\rm ff}$ is the free-fall time \citep[e.g.,][]{schmidt1959, kennicutt1998}. The term $M_{\rm gas}/t_{\rm ff}$ can be interpreted as the gas inflow rate onto the galaxy\footnote{We note that alternative definitions of the inflow rate may be adopted. However, our results remain unchanged, as the inflow-rate term cancels out in the definition of the mass-loading factor.}. If the fraction of inflowing gas that is not converted into stars is expelled through outflows, then $\dot{M}_{\rm out} = (1-\epsilon_\star) M_{\rm gas}/t_{\rm ff}$, which combined with the SFR expression yields $\eta = (1-\epsilon_\star)/\epsilon_\star$.
Using the mass-loading factors inferred from the \ciii\ emission, we infer $\epsilon_\star= 0.4^{+0.3}_{-0.2}$ in the interpretation in which the displaced gas is associated with past outflows. We regard this as a conservative estimate. In contrast, we find $\epsilon_\star= 0.08^{+0.16}_{-0.05}$ if the \ciii\ emission is interpreted as arising from an ongoing outflow.  
We note that the $\epsilon_\star$ estimates should be regarded as upper limits, as they assume that ejective feedback is the sole mechanism regulating star formation. Given the conservative assumptions adopted in our outflow-property estimates, we consider $\epsilon_\star < 8\%(^{+16}_{-5})$ as our fiducial upper limit value.

Figure~\ref{fig:sfe} shows the star-formation efficiency, $\epsilon_\star$, as a function of halo mass ($M_{\rm h}$) for different semi-analytic models used to reproduce the $z>10$ UV luminosity function. In detail, we report the density-modulated SFE model by \cite{somerville2025} for three different values of the gas fraction in dense clouds, ${\rm f_d}= 1, 0.5, 0.1$, the empirical model by \cite{yung2025} based on abundance matching. We also include the predictions from the Feeback-Free Burst \citep{dekel2023, li2024} and Attenuation Free models \citep{ferrara2023, ferrara2024a}.

The upper limits on $\epsilon_\star$ derived for the two scenarios are shown as black circles, with the corresponding error bars indicating the associated uncertainties. 
Given that the stellar mass of \gsz\ is $M_\star = 10^{8.5-9.0}~{\rm M_\odot}$ \citep{carniani2024, carniani2025, helton2025}, the inferred values of $\epsilon_\star$ also set a lower limit on $M_{\rm h}$ through the relation $M_\star/M_{\rm h} = f_b \epsilon_\star$,  where $f_b = \Omega_b/\Omega_m = 0.158$ is the cosmological baryon fraction. We therefore find that the halo mass of \gsz\ must be larger than $10^{9.4}~{\rm M_\odot}$ and $10^{9.9}~{\rm M_\odot}$ in the past- and ongoing-outflow scenarios, respectively.

In the halo mass regimes of \gsz, our fiducial constraint, $\epsilon_\star \lesssim 8\%$, disfavors models that require extreme star-formation efficiencies. Instead, it is consistent with the Attenuation-Free model and the extrapolation to high halo masses of the predictions from the density-modulated SFE models in which the fraction of gas in dense clouds is below $f_d\lesssim0.5$. Overall, these results indicate that a combination of moderate ($\sim$8\%)  star formation efficiency and reduced dust attenuation can explain the emergence of bright galaxies at the Cosmic Dawn.

%%%%%%%%%%%%%%%%%%%%%%%%%%%%%%%%%%%%%%%%%%%%%%%%%%%%%%%%%%%%%%
\section{Summary and conclusion}

We present new  NIRSpec MSA PRISM/CLEAR observations of JADES-GS-z14-0 (denoted as \gsz\ in this work) obtained with the JADES and OASIS programmes. The OASIS data reveal a \ciii\ emission line that is 2.7 times brighter than in the JADES observations, despite showing a similar rest-frame UV continuum. We summarise our main findings below.
\begin{itemize}
\item[$\blacksquare$]  Since the NIRSpec shutter assigned to \gsz\ in the OASIS programme is shifted by $\sim0.1^{\prime\prime}$ with respect to the JADES configuration, the new data suggest that the \ciii\ emission is spatially offset by $\sim400$ pc eastwards of \gsz, so explaining its weakness in  \cite{carniani2025}.

\item[$\blacksquare$]  By comparing the OASIS spectrum with that of the neighbouring galaxy 183348 at $z=3.475$, we exclude the possibility that the newly detected line is associated with H$\alpha$ emission from the lower-redshift galaxy. 

\item[$\blacksquare$] Despite its brightness, the spatially offset \ciii\ line is not detected in the deep JADES NIRCam medium-band imaging. This implies that the \ciii-emitting region has a surface brightness below the detection threshold, and therefore must be extended on scales larger than $\sim165$ pc.

\item[$\blacksquare$] Given these spatial constraints, we conclude that the \ciii\ emission originates from diffuse, carbon-enriched gas spatially offset from the stellar continuum emission of \gsz.  Such enriched material can plausibly trace either an ongoing outflow or the remnants of past outflow episodes. From the \ciii\ luminosity, we infer an ionised-gas mass of ${\rm  10^{8.5\pm0.5}~M_\odot}$. Assuming that this gas is moving with a velocity of $170~{\rm km s^{-1}}$, a conservative estimate required to reach the observed projected distance from the galaxy, implies a mass outflow rate of ${\rm 160~M_\odot~yr^{-1
}}$, yielding to a mass-loading factor between $\eta= 4$ and $\eta=15$. This value is an order of magnitude higher than typically inferred at $z\sim5$ ($\eta\sim0.04-1$; \citealt{carniani2024a, xu2025, saldana2025, parlanti2025, delpino2026, zamora2025}), suggesting the presence of powerful outflows at early epochs. 
%Compared to the H$\alpha$-based ${\rm SFR=9~M_\odot~yr^{-1 
\item[$\blacksquare$]  We also show that if outflows were the sole mechanism regulating star formation in \gsz, the implied instantaneous star-formation efficiency would be below $\sim40\%$, with a fiducial upper limit of $\sim8\%$.
\end{itemize}

Our results place constraints on galaxy-formation models by setting an upper limit on the instantaneous star-formation efficiency. Nevertheless, high spectral resolution observations with JWST in integral-field spectroscopy mode and with ALMA will be crucial to measure the line profile of optical or far-infrared transitions and map the morphology to determine the spectral and spatial displacement of the emitting gas. In addition, resolving the \ciii\ doublet would provide a density diagnostic, reducing the uncertainty on the gas mass.

More broadly, our analysis highlights that the presence (or apparent absence) of rest-frame UV emission lines may depend on the precise placement of the MSA shutter. This effect may help explain why most galaxies at $z>10$ show weak or absent UV emission lines, with only a few notable exceptions. In the scenario proposed for JADES-GS-z14-0, the lack of detected emission lines may be associated with outflows that have displaced gas to large scales, resulting in spectra without emission features when the spectroscopic observations do not cover the spatially offset emitting region. Dedicated IFS campaigns targeting bright galaxies at $z>10$ will therefore be essential to determine whether spatially offset UV line emission is common and, if so, to enable statistical constraints on outflow mass-loading factors and the efficiency of star formation in the earliest massive galaxies.

%%%%%%%%%%%%%%%%%%%%%%%%%%%%%%%%%%%%%%%%%%%%%%%%%%%%%%%%%%%%%%
\begin{acknowledgements}
      We thank Rachel Somerville and Aaron Yung  for providing their model predictions.
      SC, EP \& GV acknowledge support by European Union’s HE ERC Starting Grant No. 101040227 - WINGS. 
      The project is supported by the Cosmic Dawn Center (DAWN) that is funded by the Danish National Research Foundation under grant DNRF140. 
      RM, WB, FDE, XJ \& JS acknowledge support by the Science and Technology Facilities Council (STFC), ERC Advanced Grant 695671 “QUENCH”, and by the UKRI Frontier Research grant RISEandFALL. RM also acknowledges funding from a research professorship from the Royal Society.
      SA acknowledges grant PID2021-127718NB-I00 funded by the Spanish Ministry of Science and Innovation/State Agency of Research (MICIN/AEI/ 10.13039/501100011033).
      AJB \& JC   acknowledge funding from the “FirstGalaxies” Advanced Grant from the European Research Council (ERC) under the European Union’s Horizon 2020 research and innovation programme (Grant agreement No. 789056). 
      ECL acknowledges support of an STFC Webb Fellowship (ST/W001438/1). 
      DJE, EE, BDJ, GR, MR, FS, and CNAW are supported by JWST/NIRCam contract to the University of Arizona NAS5-02015. 
      DJE is also supported as a Simons Investigator.  BER acknowledges support from the NIRCam Science Team contract to the University of Arizona, NAS5-02015, and JWST Program 3215.
      ST acknowledges support by the Royal Society Research Grant G125142.
      H\"U thanks the Max Planck Society for support through the Lise Meitner Excellence Program. H\"U acknowledges funding by the European Union (ERC APEX, 101164796). Views and opinions expressed are however those of the authors only and do not necessarily reflect those of the European Union or the European Research Council Executive Agency. Neither the European Union nor the granting authority can be held responsible for them
      The research of CCW is supported by NOIRLab, which is managed by the Association of Universities for Research in Astronomy (AURA) under a cooperative agreement with the National Science Foundation. 
      PGP-G acknowledges support from grant PID2022-139567NB-I00 funded by Spanish Ministerio de Ciencia e Innovaci´on MCIN/AEI/10.13039/501100011033, FEDER, UE.
      J.M.H. is supported by JWST Program 8544.
\end{acknowledgements}

  \bibliographystyle{aa} % style aa.bst
  \bibliography{gsz14_ciii_biblio} % your references Yourfile.bib

%%%%%%%%%%%%%%%%%%%%%%%%%%%%%%%%%%%%%%%%%%%%%%%%%%%%%%%%%%%%%%%
% Appendices must be placed after   \end{thebibliography}
% They will be placed automatically on a new page.
%%%%%%%%%%%%%%%%%%%%%%%%%%%%%%%%%%%%%%%%%%%%%%%%%%%%%%%%%%%%%%%
%%%%%%%%%%%%%%%%%%%%%%%%%%%%%%%%%%%%%%%%%%%%%%%%%%%%%%%%%%%%%%%
\begin{appendix}
\nolinenumbers

\section{Shorts in OASIS data}

\begin{figure}
   \centering
   \includegraphics[width=\hsize]{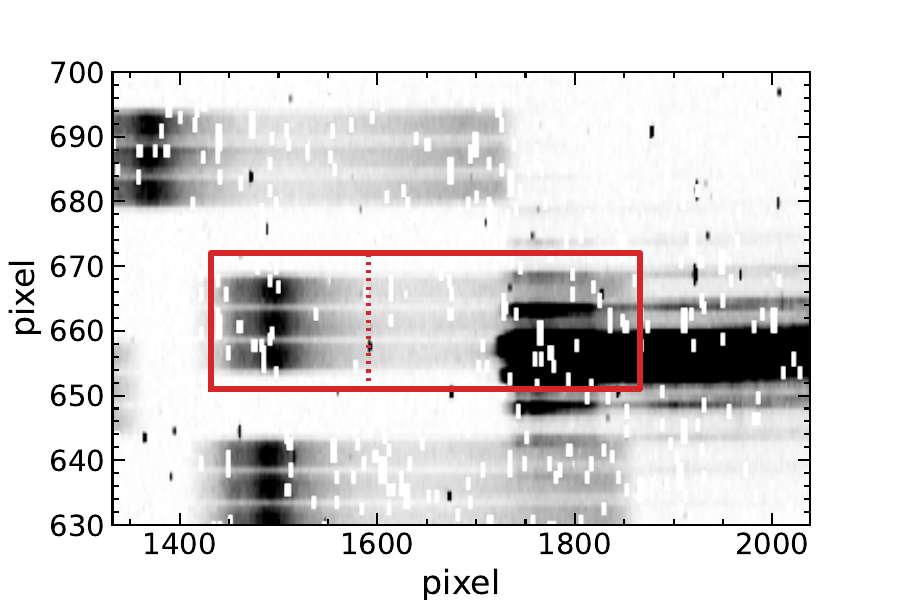}
   \caption{
   Zoom-in of the count-rate image from one of the MOS PRISM/CLEAR exposures of the OASIS programme. The red contour shows the location of the 3-shutter slitlet assigned to \gsz.  The red vertical dotted line indicates the position along the slitlet corresponding to $2.89\,\mu$m, where the \ciii\ emission line is detected.
   The glow at the end of the trace is the artefact signal caused by the electronic shorts. The part of the count-rate image contaminated by the artificial glow is the origin of the altered spectrum profile at wavelengths longer than 4.2\,$\mu$m.}
         \label{fig:shorts}
   \end{figure}

Figure~\ref{fig:shorts} shows a zoom-in of the count-rate image from one of the MOS PRISM/CLEAR exposures obtained as part of the OASIS programme, highlighting the diffuse glow caused by electronic shorts in the microshutter array. The position of the 3-shutter slitlet assigned to \gsz\ and used for the spectral extraction is outlined by a red contour. The contamination associated with this glow affects the portion of the spectrum at wavelengths longer than $4.3\,\mu$m, producing an artificial signal that remains visible even after background subtraction using the standard nodding scheme (see Fig.~\ref{fig:spectra}). The Figure also illustrates that the region corresponding to the detected \ciii\ emission at $2.89~\mu$m, marked by a vertical dotted line in the count-rate map, is not affected by the shorts.

%-------------------------------------------------------------
%-------------------------------------------------------------

\section{\ciii emission in NIRCam}

\begin{figure}
\centering
	\includegraphics[width=0.9\columnwidth]{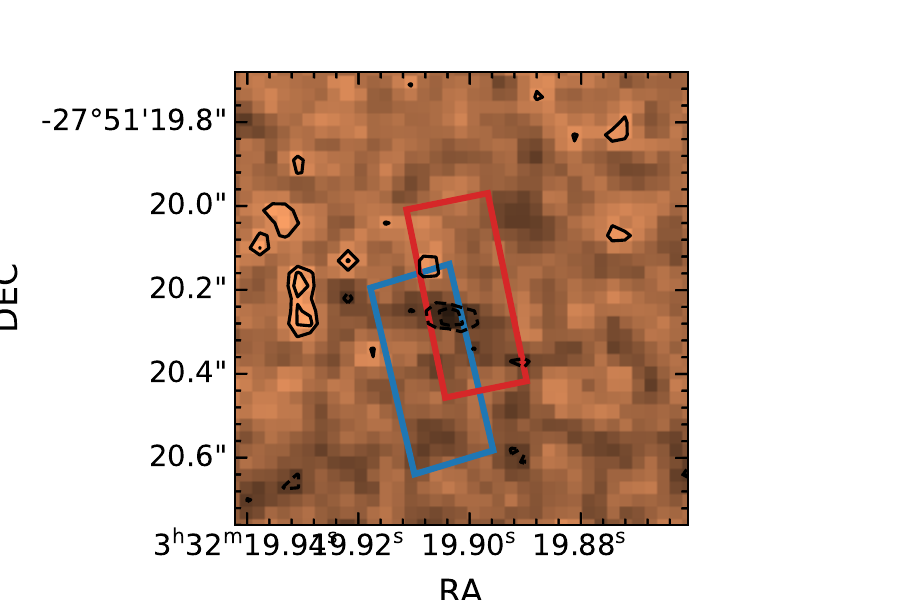}
    \caption{NIRCam F300M image after subtraction of the NIRCam F250M image. Contours correspond to $-3\sigma$, $-2\sigma$, $2\sigma$, and $3\sigma$. The slit positions of the OASIS and JADES programmes are shown in blue and red, respectively.}
    \label{fig:res}
\end{figure}

Figure~\ref{fig:res} shows the image \gsz obtained by subtracting the F300M image from the F250M image.  The residual image has a $5\sigma$ point-source sensitivity of 2.6~nJy. We do not identify any clear emission at the location of the OASIS shutter that can be associated with the \ciii emission suggesting the emission is spatially resolved (see Section~\ref{sec:offset})

%-------------------------------------------------------------

\section{Photoionization from the galaxy}\label{sec:photoionization}

In the outflow scenario, we assume that the displaced gas cloud is photoionised by radiation emitted by \gsz. To test this hypothesis, we use the \textsc{CLOUDY} code to estimate the expected \ciii\ luminosity from a cloud illuminated by a stellar population spectrum with a total luminosity comparable to that inferred for \gsz\ ($\sim10^{45.3}~{\rm erg\,s^{-1}}$), varying the distance of the cloud and comparing the predictions with the observations. 

\begin{figure}
\centering
	\includegraphics[width=0.9\columnwidth]{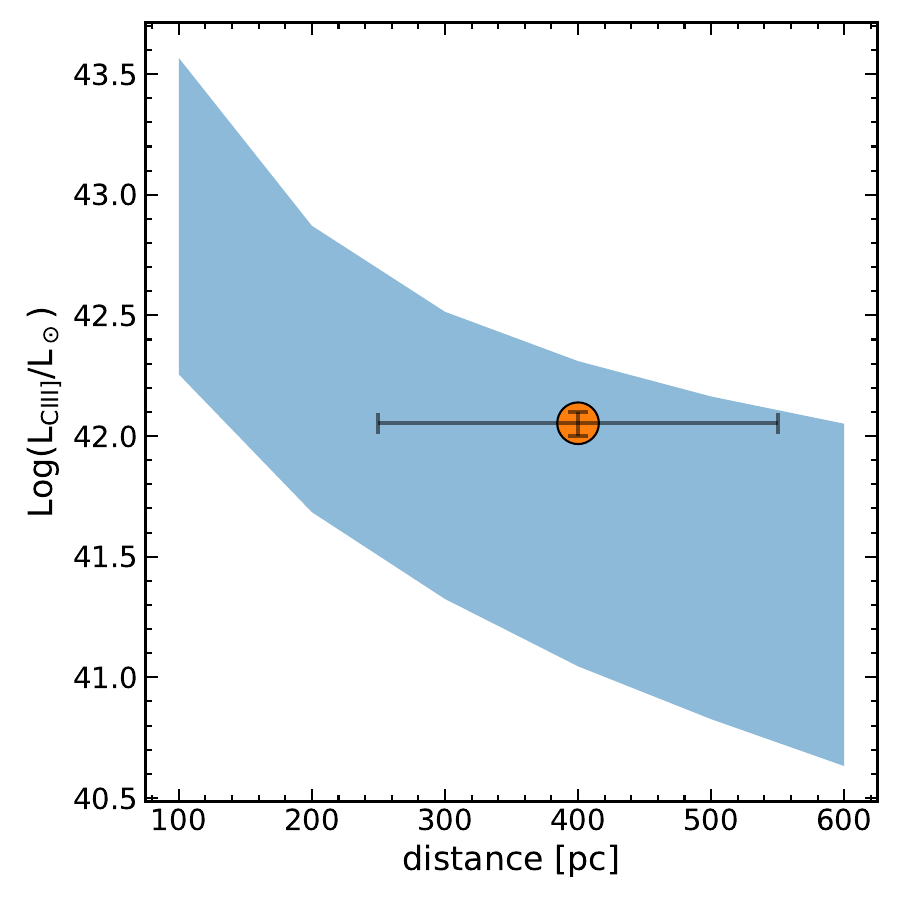}
    \caption{Expected \ciii luminosity resulting from a cloud photoionized by a stellar population with a luminosity as bright as \gsz, as a function of distance from the latter. The yellow circle shows the \ciii luminosity measured in OASIS and the distance of the gas cloud based on the position of the NIRSpec MOS shutters.}
    \label{fig:photoionization}
\end{figure}

We assume a spherical geometry for the cloud, with a radius of 165 pc, corresponding to the lower limit derived from the analysis of the NIRCam imaging (see Sect.~\ref{sec:offset}). We compute the \ciii\ luminosity as a function of distance for a range of electron densities ($10^{1.5}~{\rm cm^{-3}} < n_e < 10^{3.5}~{\rm cm^{-3}}$) and gas-phase metallicities ($0.1~{\rm Z_\odot} < Z < 0.5~{\rm Z_\odot}$). Finally, we adopt synthetic spectra generated with \textsc{Bagpipes} as the ionising source, exploring stellar population ages in the range $1~{\rm Myr} < \tau_\star < 20~{\rm Myr}$. The light blue shaded region in Figure~\ref{fig:photoionization} shows the expected \ciii luminosity of the ionized cloud as a function of distance from \gsz, while the yellow symbol shows the measured \ciii\ luminosity in OASIS data. Our results indicate that a gas cloud located at a distance of $\sim300-500$ pc from \gsz\ can reproduce the observed \ciii\ emission if externally excited by ionising photons escaping from the galaxy.

\end{appendix}
\end{document}